\documentclass{elsart}
\usepackage{colordvi}
\usepackage{color}
\usepackage{epsfig}

\begin{document}

\begin{frontmatter}
\title{
Droplet formation in expanding nuclear matter: \\ 
a system-size dependent study} 

\author[GSI]{W.~Reisdorf},
\author[STRAS]{F.\,Rami},
\author[STRAS]{B.\,de Schauenburg},
\author[GSI]{Y.~Leifels},
\author[CLER]{J.P.\,Alard},
\author[GSI]{A.~Andronic},
\author[CLER]{V.~Barret},
\author[ZAG]{Z.\,Basrak},
\author[CLER]{N.\,Bastid},
\author[HEID]{M.L.\,Benabderrahmane},
\author[ZAG]{R.\,\v{C}aplar},
\author[CLER]{P.~Crochet},
\author[CLER]{P.\,Dupieux},
\author[ZAG]{M.\,D\v{z}elalija},
\author[BUD]{Z.\,Fodor},
\author[GSI]{A.~Gobbi},
\author[ITEP]{Y.\,Grishkin},
\author[GSI]{O.~N.~Hartmann},
\author[HEID]{N.~Herrmann},
\author[GSI]{K.D.~Hildenbrand},
\author[KOR]{B.\,Hong},
\author[BUD]{J.\,Kecskemeti},
\author[KOR,GSI]{Y.J.~Kim},
\author[GSI,HEID,WAR]{M.~Kirejczyk},
\author[GSI]{P.~Koczo\'{n}},
\author[ZAG]{M.\,Korolija},
\author[ROSS]{R.\,Kotte},
\author[GSI]{T.~Kress},
\author[ITEP]{A.\,Lebedev},
\author[CLER]{X.~Lopez},
\author[HEID]{M.\,Merschmeyer},
\author[ROSS]{J.\,M\"{o}sner},
\author[ROSS]{W.\,Neubert},
\author[HEID]{D.\,Pelte},
\author[BUC]{M.\,Petrovici},
\author[GSI]{A.~Sch\"{u}ttauf},
\author[BUD]{Z.\,Seres},
\author[WAR]{B.\,Sikora},
\author[KOR]{K.S.\,Sim},
\author[BUC]{V.\,Simion},
\author[WAR]{K.\,Siwek-Wilczy\'nska},
\author[ITEP]{V.\,Smolyankin},
\author[HEID]{M.\,Stockmeier},
\author[BUC]{G.\,Stoicea},
\author[GSI,WAR]{Z.~Tymi\'{n}ski},
\author[STRAS]{P.\,Wagner},
\author[WAR]{K.~Wi\'{s}niewski},
\author[ROSS]{D.\,Wohlfarth},
\author[GSI]{Z.G.~Xiao},
\author[KUR]{I.\,Yushmanov},
\author[ITEP]{A.\,Zhilin}

(FOPI Collaboration)
\address[GSI]{Gesellschaft f\"ur Schwerionenforschung, Darmstadt, Germany}
\address[STRAS]{Institut de Recherches Subatomiques, IN2P3-CNRS, Universit\'e
Louis Pasteur, Strasbourg, France}
\address[CLER]{Laboratoire de Physique Corpusculaire, IN2P3/CNRS,
and Universit\'{e} Blaise Pascal, Clermont-Ferrand, France}
\address[ZAG]{Rudjer Boskovic Institute, Zagreb, Croatia}
\address[HEID]{Physikalisches Institut der Universit\"at Heidelberg,
Heidelberg, Germany}
\address[BUD]{Central Research Institute for Physics, Budapest, Hungary}
\address[ITEP]{Institute for Theoretical and Experimental Physics, Moscow,
Russia}
\address[KOR]{Korea University, Seoul, South Korea}
\address[WAR]{Institute of Experimental Physics, Warsaw University, Poland}
\address[ROSS]{Forschungszentrum Rossendorf, Dresden, Germany}
\address[BUC]{National Institute for Nuclear Physics and Engineering, Bucharest,
Romania}
\address[KUR]{Kurchatov Institute, Moscow, Russia}



\begin{abstract}
Cluster production is investigated
in central collisions of Ca+Ca, Ni+Ni, $^{96}$Zr+$^{96}$Zr, 
$^{96}$Ru+$^{96}$Ru, Xe+CsI and Au+Au
reactions at  $0.4A~{\rm GeV}$ incident energy.
We find that the multiplicity of  clusters with charge $Z \ge 3$ 
grows quadratically with the system's total charge
and is associated with a mid-rapidity source
with increasing transverse velocity fluctuations.
When reduced to the same number of available charges,
an increase of cluster production by about a factor of 5.5 is observed
in the mid-rapidity region
between the lightest system (Ca+Ca) and the heaviest one (Au+Au).
The results, as well as simulations using Quantum Molecular Dynamics,
suggest a collision process where droplets,
{\rm{i.e.}} nucleon clusters, are created in an expanding,
gradually cooling, nucleon gas.
Within this picture, expansion dynamics, collective radial flow
and cluster formation are closely linked as a result of
the combined action of nucleon-nucleon scatterings and the mean
fields.
\end{abstract}

\end{frontmatter}

In energetic central heavy ion collisions it is generally assumed that, after
going through an early stage of hot and compressed nuclear matter, the system
undergoes a substantial expansion causing local
cooling {\em before} freezing out.
At beam energies  above 10A GeV the current picture is that the hot system
while cooling passes from a phase involving at least partially deconfined
quarks and gluons into a purely confined hadronic phase.
At energies below 1A GeV the hot phase is still predominantly a nucleonic gas
which, however, in the expansion phase can partially 'liquefy' i.e. clusterize
in analogy to the processes used
in clusterization devices for atomic physics~\cite{Heer93}, although
under less controlled conditions.
In both energy regimes the aim is  to determine basic parameters, such
as the critical temperature $T_c$ or the latent heat of the (first order) phase
transition.
Due to the finite size of the nuclear systems available in accelerator physics
and the complexity of the dynamics of heavy ion reactions, convincing progress
on this frontier has proven to be a difficult task.
Concerning the liquid-to-gas transition onsets of plateau's in caloric
curves~\cite{pochodzalla95} have been interpreted~\cite{natowitz02}
in ways that relate 
indirectly to first order transitions.
More direct signatures~\cite{chomaz02,borderie02},
such as
negative heat capacities~\cite{dagostino00} have been subjected to critical
reviewing~\cite{moretto02}.
Claims to the determination of
$T_c$~\cite{elliott02,kleine02,karnaukhov03} vary  in the proposed values 
and require model assumptions, that seem to be in conflict with
some experimental data~\cite{williams97}.
Indications for the existence of a negative compressibility region
in the nuclear phase diagram, leading to spinodal instabilities
characterized by enhancement of events with nearly equal-sized fragments,
have been obtained~\cite{borderie01}.
The very small cross sections for this phenomenon were justified
with microscopic simulations.
One of the key assumptions in many works is that multifragmentation is a unique
mechanism related directly to subcritical and/or critical
phenomena~\cite{borderie02}.
In recent theoretical simulations using Nuclear Molecular Dynamics (NMD),
together with a backtracing method,
Bondorf et al.~\cite{bondorf95} have argued that two mechanisms
coexist:
a dynamical  rupture of spectator-type, relatively cold, fragments
and a second process where nucleon-nucleon collisions generate the seeds for
completely new fragment creation, coalescing nucleons 
{\em which were initially far from each other in phase space}.

In this Letter we present data for central collisions in symmetric heavy ion
systems at 0.4A GeV that strongly support these theoretical
findings~\cite{bondorf95}.
In order to better understand the finite-size problem, we have varied
the system size from Ca+Ca ($Z_{sys}=40$)  to Au+Au ($Z_{sys}=158$),
investigating five systems of different size.
We find that the multiplicity of heavy clusters with charge $Z \ge 3$,
when reduced to the same number of available charges, grows linearly
with system size and is associated with a mid-rapidity source
with increasing transverse momentum fluctuations.
This is in strong contrast to multifragmentation of quasi-projectiles in the
Fermi energy regime or of spectator matter at SIS/BEVALAC 
energies (0.1-2A GeV) where a high degree of 
'universality'~\cite{schuettauf96,beaulieu96} was observed, with among others
an apparent system-size independence.
As parallel studies of the same systems~\cite{reisdorf03} have shown an
increasing degree of stopping as well as increasing generation of flow
when passing from light to heavy systems,
we  can conclude that the increased flow
leads to a cooling process favouring gradual 'liquefaction'.
This is a non-trivial finding:
from earlier experimental studies~\cite{kunde95} a suppression of
heavy fragment production in systems with strong collective expansion was
inferred and theoretical works~\cite{chikazumi01}
have  predicted that strong flow gradients would prevent coagulation.
A similar behavior is seen in
two-particle correlations where the effective radii of homogeneity are
diminished by flow~\cite{heinz99}.

 The data were taken 
 at the SIS accelerator of {\small GSI}-Darmstadt 
 using various heavy ion beams of $0.4A$ GeV and the large acceptance
 {\small FOPI} detector~\cite{gobbi93,ritman95}.
In the experiments involving the systems 
$^{40}$Ca+$^{40}$Ca,                                    
$^{96}$Ru+$^{96}$Ru, $^{96}$Zr+$^{96}$Zr,  and     
$^{197}$Au+$^{197}$Au,
particle tracking  and energy loss determinations were
done using two drift chambers, the CDC (covering laboratory polar angles
between $35^\circ$ and $135^\circ$) and the Helitron ($9^\circ-26^\circ$), both
located inside a superconducting solenoid operated at a magnetic field of
0.6T.
A set of scintillator arrays, Plastic Wall $(7-30)^0$, Zero Degree Detector
$(1.4^\circ-7^\circ)$, and Barrel $(42^\circ-120^\circ)$, allowed
to measure the time of flight
and, below $30^\circ$, also the energy loss.
All subdetector systems have full azimuthal coverage.
Use of CDC and Helitron allowed the identification of pions, as well as
good isotope separation for  hydrogen and
helium clusters in a large part of momentum space.
The identification of 
heavier clusters ($Z\ge 3$), by nuclear charge only, was restricted to the
polar angles covered by the Plastic Wall and the Zero degree detector.
In a second setup the Helitron was replaced by
an array of gas ionization chambers~\cite{gobbi93}, the PARABOLA,
allowing charge identification of heavier clusters up to nuclear charge $Z=12$.
The systems
$^{58}$Ni+$^{58}$Ni,                          
$^{129}$Xe+CsI,
$^{197}$Au+$^{197}$Au 
were studied in this experiment.
The data for Au on Au, measured with both setups, were found to be in
excellent agreement. 
Further details on the detector resolution and performance can be found 
in~\cite{gobbi93,ritman95}.

Collision centrality selection was obtained by binning
distributions of the ratio, $Erat$~\cite{reisdorf97}, of total transverse
and longitudinal kinetic energies.
In terms of the scaled impact
parameter, $b^{(0)}=b/b_{max}$, we choose the same centrality for all the
systems: $b^{(0)}<0.15$.
We take $b_{max} = 1.15 (A_{P}^{1/3} + A_{T}^{1/3})$ 
as effective sharp radius and estimate $b$
from the measured differential cross sections for the $Erat$
distribution using a geometrical sharp-cut approximation.
In this energy regime
 $Erat$ selections show better impact parameter resolution for the most
central collisions than  multiplicity selections~\cite{reisdorf97,jeong94}
and do not imply {\em a priori} a chemical bias.
Autocorrelations in high transverse momentum population, that are caused by
the selection of high $Erat$ values, are avoided by not including the
particle of interest in the selection criterion.

The analysis of the particle spectra involves some interpolations and
extrapolations ($30\%$ in the worst case) to fill the gaps in the measured
data.
The two-dimensional method used to achieve this has been extensively
tested with theoretical data generated with the
transport code IQMD~\cite{hartnack98} applying apparatus filters.
Details  will be published elsewhere~\cite{reisdorf04}.

Since the present study is restricted to central collisions
of symmetric systems, we require
reflection symmetry in the center of momentum ($c.o.m.$) and,
use azimuthally averaged data.
Choosing the $c.o.m.$ as reference frame and orienting the z-axis in the beam
direction the two remaining dimensions are characterized by the longitudinal
rapidity $y\equiv y_z$,
given by $exp(2y)=(1+\beta_z)/(1-\beta_z)$ and the transverse
component of the four-velocity $u$, given by $u_t=\beta_t\gamma$.
Following common notation $\vec{\beta}$ is the velocity in units of the light
velocity and $\gamma=1/\sqrt(1-\beta^2)$. 
Later we shall also use the {\em transverse} rapidity, $y_x$, 
which is defined by
replacing $\beta_z$ by $\beta_x$ in the expression for the longitudinal
rapidity. 
The $x$-axis is laboratory fixed and hence randomly oriented relative to the
reaction plane, i.e. we average over deviations from axial symmetry.
Throughout we use scaled units $y^{(0)}=y/y_p$ and $u_{t}^{(0)}=u_t/u_p$,
with $u_p=\beta_p \gamma_p$, the index {\em p} referring to the incident
projectile in the $c.o.m.$.

\begin{figure}[t]
\centering{\mbox{\hspace{-1.0cm}
\epsfig{file=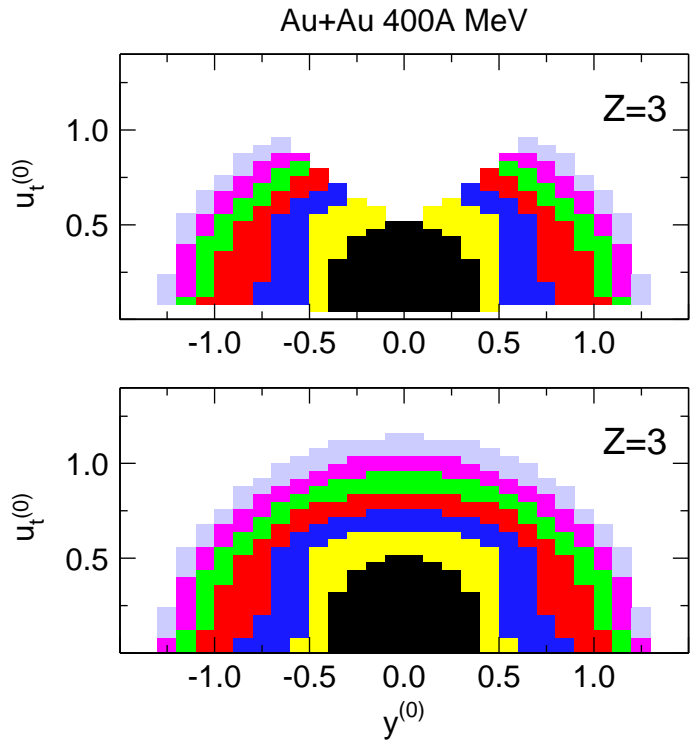,width=12cm}}}
%
\caption{
Invariant distribution $dN/(u_{t}^{(0)} du_{t}^{(0)} dy^{(0)})$ 
of Li ions emitted in central collisions of Au+Au at $0.4A$ GeV.
The various grey (color) tones correspond to yields differering by a factor
of 1.5. 
The upper panel shows the measured data                       
including the use of reflection symmetry and interpolations with the
two-dimensional fit method.
The lower panel shows the extension to $4\pi$.
}
\label{f:fig1}
\end{figure}

An example of a reconstructed distribution for emitted Li ions in central
collisions of Au on Au is shown in Fig.~\ref{f:fig1}. The
lower part of this figure illustrates the result of the extrapolation to
$4\pi$ using the two-dimensional fit method.

\begin{figure}[htb]
\begin{center}
\epsfig{file=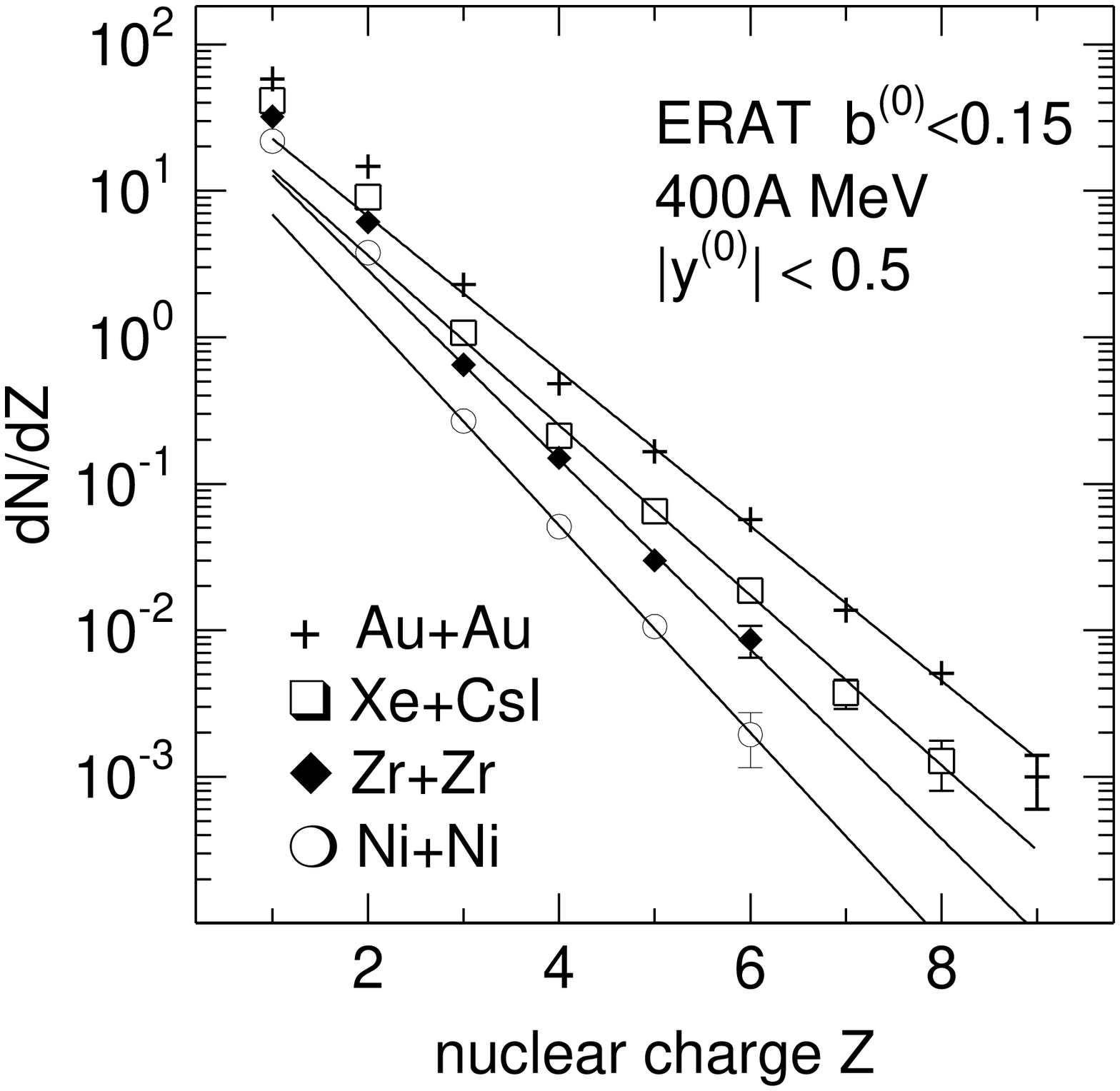,width=13cm}
\end{center}
\caption{
Charged particle multiplicity distributions 
 - $dN/dZ$ - measured in central collisions ($b^{(0)} < 0.15$) 
of Au+Au, Xe+CsI, Zr+Zr and Ni+Ni reactions at an
incident energy of $0.4A~{\rm GeV}$.
The data are presented for {\em c.o.m.}
rapidities $ \colon |{y^{(0)}}| \le 0.5$.
The straight lines are exponential fits to the data in the interval
$3\leq Z \leq 6$.
Errors, if not shown, are smaller than the symbol sizes.
}
\label{f:fig2}
\end{figure}

Turning now to the presentation of results, we show in Fig.~\ref{f:fig2}
charged particle multiplicity distributions as a function
of nuclear charge $Z$.
The data for four out of the six measured systems are plotted.
As the surface of nuclei has a finite thickness one expects some degree of
transparency even in collisions with perfect geometrical overlap.
To minimize such 'corona' effects, we show midrapidity data, $|y^{(0)}| < 0.5$. 
As observed earlier~\cite{reisdorf97},
we find that the heavy cluster ($Z > 2$)  multiplicities, $M_{hc}$,
decrease exponentially with the nuclear charge, i.e. $M_{hc}(Z) \sim
\exp(-c_{hc}^{*}Z)$, however, the slope parameter $c^*_{hc}$ 
is seen to vary with the
system-size (see Fig.~\ref{f:fig2} and Table 1):
{\em heavy cluster production is favoured for larger systems}.

In the present context  we refrain from calling heavy clusters (hc), $Z > 2$,
'intermediate mass fragments'
 (IMF), because in this energy regime they are, actually, the {\em
heaviest} fragments ('droplets') in central collisions.
Charge balances show unambiguously  that there is no heavy
remnant ($Z>1/6 Z_{sys}$) 
with a sizeable ($>1\%$) probability (that one might be tempted to call
'liquid').
The slope parameters, obtained for the range $Z=3-6$,
are listed in Table 1 both for the mid-rapidity
data, $c^*_{hc}$, and the $4\pi$ data, $c_{hc}$.
For Ca+Ca reliable data beyond $Z=4$ could not be obtained.
The value $c_{hc}=1.224\pm0.043$ for Au+Au can be compared
with our earlier~\cite{reisdorf97} value of $1.170\pm0.018$ 
which was obtained from a fit over a larger $Z$ range.
Although a different method to extrapolate to $4\pi$ was used,
the main reason for the modest deviation can be traced to the
somewhat higher centrality achieved with the present setup compared with the
'PHASE I' setup used in our earlier work, which covered only polar angles below
$30^{\circ}$. 

If one were to interprete the slopes as being an indicator of a global   
freeze-out temperature of systems in chemical equilibrium,
the qualitative conclusion would be that
{\em smaller systems appear to be hotter}.
We recall that all systems are studied at the same incident energy and for
the same centrality, $b^{(0)} = b/b_{max} < 0.15$.
However,
in complex reactions leading to many outgoing channels with an apparently
simple (exponential)
statistical distribution, integrated charge distributions 
give limited information 
and therefore do not allow to draw convincing conclusions on the 
possible emergence of a final state {\em in equilibrium}.


\begin{figure}[htb]
\begin{center}

\epsfig{file=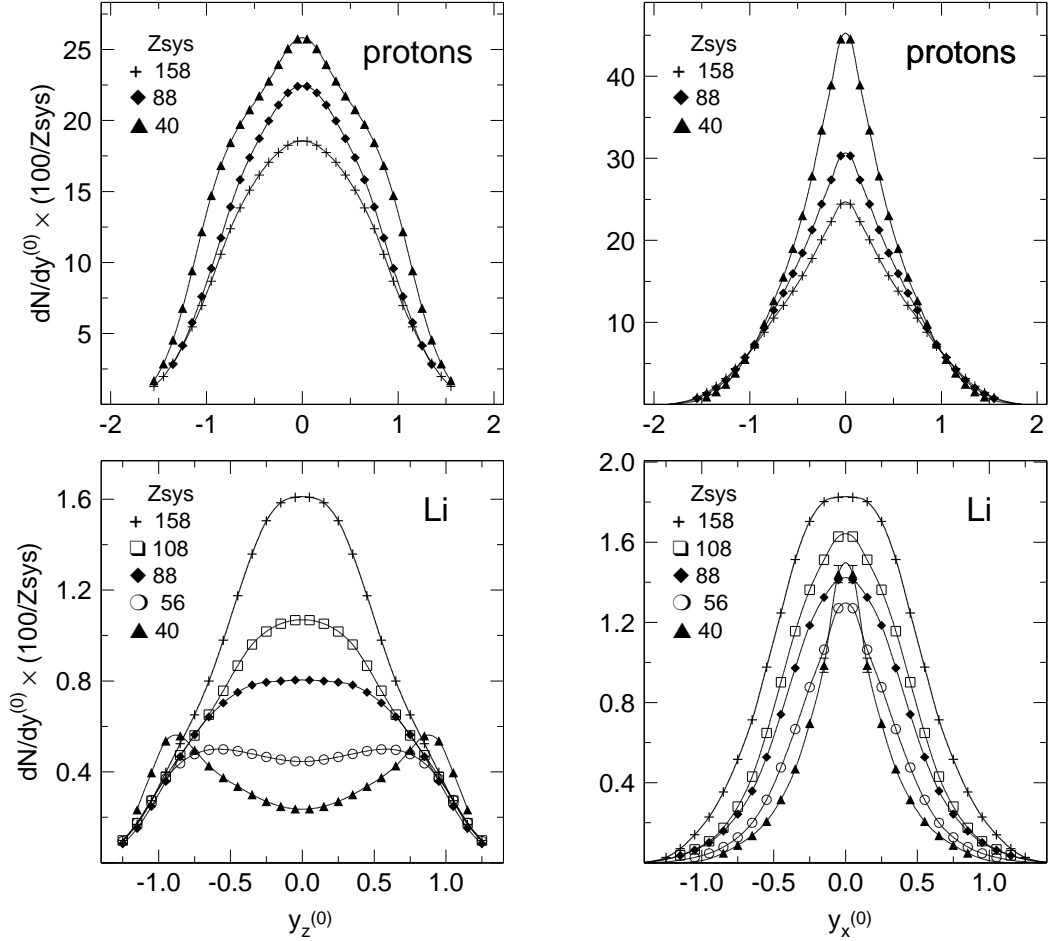,width=15cm}
\end{center}
\caption{
Scaled longitudinal (left) and transverse (right) rapidity distributions
of single protons (top) and Li clusters (bottom) measured in central
collisions for different symmetric reactions. 
The various system charges $Z_{sys}$ are indicated in the Figure.
The ordinates are normalized to a common reference system with charge
$Z_{sys}=100$.
}
\label{f:fig3}
\end{figure}

\begin{table}[htb]
\caption{Characteristics of heavy clusters production in central heavy ion
collisions at $0.4A$ GeV.}
\label{tab1}
\begin{center}
\begin{tabular}{llllll}
\hline
System & $c_{hc}$ & $c_{hc}^*$ & $M_{hc}$ & $M_{hc}^*$ & {\em vart(Li)} \\
       &     &    & $\pm 10 \%$ & $\pm 10\%$   &  $\pm 0.02$     \\
\hline
$^{40}$Ca + $^{40}$Ca &    &   & 0.42 & 0.133 & 0.094  \\
$^{56}$Ni + $^{40}$Ni &  1.44 $\pm$ 0.07 & 1.63 $\pm$ 0.08 
& 0.75 & 0.32 & 0.117  \\ 
$^{96}$Zr + $^{96}$Zr & 1.441 $\pm$ 0.046  & 1.49 $\pm$ 0.06
 & 1.55 & 0.84 & 0.139  \\
$^{96}$Ru + $^{96}$Ru &  1.415 $\pm$ 0.042 & 1.37 $\pm$ 0.06 & 1.65 &
 0.89  & 0.144  \\
$^{108}$Xe + $^{133}$Cs$^{127}$I & 1.276 $\pm$ 0.040 & 1.336 $\pm$ 0.044 
       & 2.29 & 1.35 & 0.147  \\
$^{197}$Au + $^{197}$Au & 1.224 $\pm$ 0.043 & 1.217 $\pm$ 0.044 & 4.66
       & 2.99 & 0.177 \\
\hline
\end{tabular}
\end{center}
\end{table}


More details of the mechanism at work are seen in Fig.~\ref{f:fig3}
where we  compare the
evolution of the longitudinal and the transverse rapidity distributions
(left and right panels, respectively)
with system size for the 'gas' (protons) which is prevalent for the
light (surface-dominated) systems, and the 'droplets' which appear in
increasing numbers at {\em mid-rapidity} as the system-size increases 
(lower left panel),
accompanied by a broadening of the transverse velocities (lower right panel).
We plot the smoothened data emphasizing the gradual evolution with
system size.
The statistical errors are smaller than the symbol sizes, while the absolute
systematic errors are $10\%$.
Relative errors are expected to be smaller by a factor of two or more.

Note that all yields in Fig.~\ref{f:fig3} are reduced to a constant
system size of 100 nuclear charges by multiplying with $(100/Z_{sys})$,
where $Z_{sys}$ is the total system charge.
Closer inspection of the system-size evolutions in these reduced scales
reveals two other noteworthy features:
1) The 'missing' proton vapour in the heaviest system, that served as source of
the developping clusters, is limited to smaller transverse rapidities,
$|y^{(0)}_x| \leq 0.7$, see the upper right panel and note the enlarged
abscissa scale in the lower right panel.
2) Except for the Ca+Ca system, heavy cluster production is system-size {\em
independent} around longitudinal rapidities, $|y^{0}| \geq 0.8$,
see the lower left panel.
This latter point is not trivial, although 'universal' features in the
partition of excited spectator matter are well established~\cite{schuettauf96}.
In general, however,the collisions investigated in ref.~\cite{schuettauf96}
were more peripheral and 'spectators' could be clearly identified as
relatively narrow peaks in the longitudinal rapidity distributions near
projectile (or target) rapidity.
In the present cases of high centrality, spectator matter, if any,
is not readily isolated in the measured rapidity distribution, except
for the two-peak structures in the Ca+Ca reaction.
But even in that case the evolution towards mid-rapidity is  continuous.
When we select more peripheral collisions 'spectator' peaks appear in all
systems and we observe~\cite{deschauen99,reisdorf99}
at this energy the same universal features as in ref.~\cite{schuettauf96}.

\begin{figure}[htb]   
\centering{\mbox{\hspace{-1.0cm}
\epsfig{file=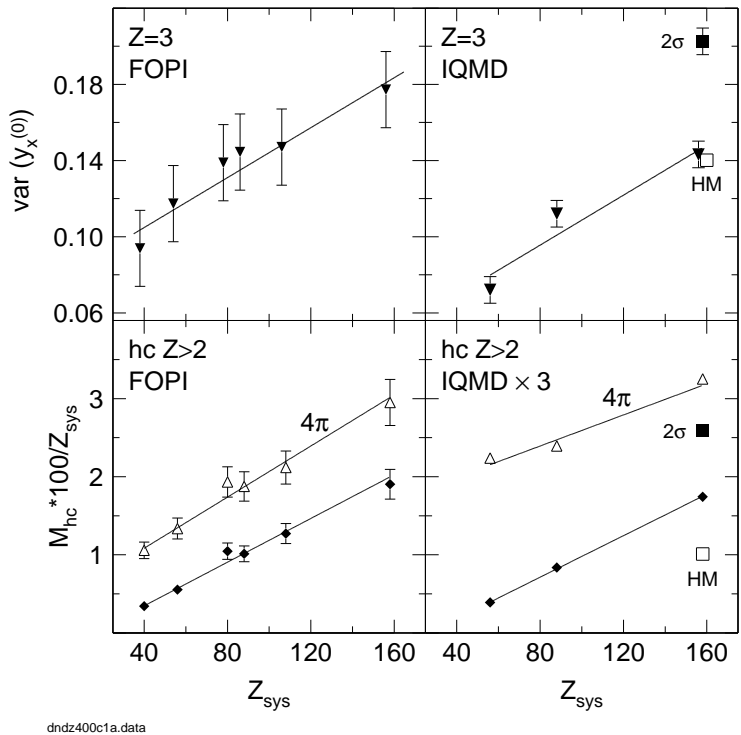,width=12.0cm}}}
\caption{
Summary of heavy cluster data measured with FOPI (left panels)
and calculated using IQMD (right panels).
Lower panels:
Average multiplicities of  heavy clusters
per 100 incoming protons ({\rm i.e.} ${{M_{hc}} \times 100 /{Z_{sys}}} $)
as a function of the system charge $Z_{sys}$.             
In decreasing order
the data represent rapidity-integrated ($4\pi$) data (open
triangles), data confined to the midrapidity interval $|y^{(0)}|<0.5$
(full diamonds) and, in the FOPI case,
data limited to the acceptance of the apparatus,
i.e before acceptance corrections (open circles).
Upper panels:
Variance of the transverse rapidity distributions for Li ions
versus system charge.
All straight lines  are linear least squares fits to the
respective data.
Note that the IQMD multiplicities are multiplied by a factor three.
The meaning of the marked square symbols in the IQMD data is explained in the
text.
}
\label{f:fig4}  
\end{figure}

A way to summarize the observations and at the same time confronting them
with theoretical simulations is presented in Fig.~\ref{f:fig4}.
Starting with the lower left panel,
we show the system-size dependence of the rapidity-integrated
multiplicity, $M_{hc}$, of heavy clusters.
Although we plot again {\em reduced} yields (to 100 incoming protons),
we
observe a remarkable linear increase: the least squares fitted straight line 
follows the data with an accuracy of $4\%$ 
(as mentioned earlier, the relative accuracy of the
data points is expected to be better than the indicated $10\%$ systematic error
bars). 
This means that the system-size dependence of $M_{hc}$ has a term proportional
to $Z_{sys}^2$ or $A_{sys}^2$ (the system mass squared).

There is a small irregularity associated with the two data points 
due to Zr+Zr ($Z_{sys}=80$), resp. Ru+Ru ($Z_{sys}=88$) which represents
systems with the same total mass ($A_{sys}=192$), suggesting that besides the
size dependence there is also an isospin dependence of heavy cluster production.
The effect is however at the limit of the error margins and therefore will not
be discussed any further in the present work. 

Besides the $A_{sys}^2$ term in $M_{hc}(A_{sys})$, a second piece of
information comes from looking at  multiplicities, $M_{hc}^*$, confined to
the midrapidity interval $(|y^{(0)}|<0.5)$ (see 
lower left panel of Fig.~\ref{f:fig4}):
the restricted data run parallel to the integrated data, showing that the
quadratic term is associated virtually entirely with the midrapidity region.
For comparison  we also show the same midrapidity data before acceptance
corrections.
While the corrections are important, and enhance the effect, it is clear that
they are not {\em creating} the effect.
In the midrapidity region the reduced cluster production is about a factor
5.5 higher for Au+Au than for Ca+Ca. 
We note that for the lightest system, $Z_{sys}=40$ is still large compared
to the most abundant $(Z=3)$ heavy cluster.
Of course the linear trends cannot go on indefinitely as the number of
heavy clusters per hundred protons cannot exceed 33 by definition.
We are far from this trivial limit, however.

We have also reduced $M_{hc}^*$ by dividing by the sum of charges accumulated
at midrapidity ($|y^{(0)} < 0.5$), $Z_{midy}$, rather than by the total system
charge, $Z_{sys}$.
In this case the straight line fitted to the data 
(not plotted to avoid overloading the Figure) 
passes very close to the origin of the axes:
the relationship $M_{hc}^* \times 100/Z_{midy}= 0.0342 Z_{midy}$ reproduces
the six data points with an accuracy of $6\%$.
This means that, well within our systematic errors, we can say that {\em all}
the mid-rapidity heavy cluster production rises with the square of the
number of 'participant' nucleon number.

The associated observation of increasingly broad {\em transverse} rapidity
distributions as the system size is increased, shown in Fig.~\ref{f:fig3},
is summarized in the upper left panel of Fig.~\ref{f:fig4} where
the variances $var(y_x^{(0)})$ of these distributions
(taken for data within $|y_x^{(0)}| < 1$) are 
plotted and also seen to rise linearly with the system size, although with
an offset.
The observables $M_{hc}$, $M_{hc}^*$ and $var(y_x^{(0)})$ are also listed in
Table 1.
In a global thermal equilibrium picture the variance of rapidity (velocity)
distributions is a measure of the temperature and hence, in this scenario, one
would conclude that {\em the heaviest system is the hottest system}, in
contradiction to our earlier conjecture, inferred from the observed partitions,
that the {\em lightest} system is the hottest.
A way out of this contradiction is to introduce radial flow: generated
during the expansion, it is accompanied by a {\em local} cooling.
This mechanism is seen to be stronger for bigger systems.
Radial flow, assessed by modelling hydrodynamic expansion~\cite{petrovici95},
or treated phenomenologically~\cite{reisdorf97}, has been introduced by our
Collaboration with some success in describing the deviations of momentum space
distributions from a {\em global} thermal scenario, notably the fragment mass
dependences.
But some inconsistency in
accounting for the cluster yields~\cite{reisdorf97} remained.
This difficulty could be due to the failure of correctly accounting for
non-equilibrium effects caused by the coronae of the nuclei and, more
generally, by partial transparency~\cite{reisdorf03}.

In principle,
such non-equilibrium effects can  be handled by simulation
codes based on transport theory~\cite{bertsch88}.
We have used the code IQMD~\cite{hartnack98} based on Quantum Molecular
Dynamics~\cite{aichelin91}, QMD, to see if we could reproduce
the features of our data.
Clusters are identified after a reaction time of 200 fm/c using the
minimum spanning tree algorithm in configuration space with a coalescence
radius of 3 fm.
For each system-energy 
we have generated 50000 IQMD events over the complete 
impact parameter range which were
subsequently sorted using the $Erat$ criterion to obtain an event class
with comparable centrality to that selected in the experiment.
The reduced heavy cluster multiplicities ($4\pi$ integrated or at           
mid-rapidity) versus system size are plotted in the lower right panel of
Fig.~\ref{f:fig4}.
As in our earlier studies~\cite{reisdorf97}, we find that IQMD,
as well as other realizations of QMD~\cite{tsang93}, underestimates
cluster production: the yields in the Figure are multiplied by a 
factor of three.
However, qualitatively, the experimental size dependence is reproduced,
including the dominant effect of mid-rapidity emissions in accounting for the
linear rise.
The associated rise of the variance of the transverse rapidity with system size
is reproduced almost quantitatively.
In this context it is useful to realize that the nuclear stopping phenomenon,
as quantized by the {\em shape} of proton and deuteron longitudinal rapidity
distributions is rather well described by IQMD~\cite{hong02} at incident
energies around $400A$ MeV.
For heavy clusters the simulations also reproduce the yields near projectile
(target) rapidity: the underestimation concerns mid-rapidity emissions and
might be a general deficiency of semiclassical approaches such as IQMD
that tend to converge towards Boltzmann statistics after many single nucleon
collisions (leading to mid-rapidity population) rather than conserving the
initial Fermi-destribution of nucleons.

To shed some light into the cluster creation mechanism, we have  performed,
for Au+Au, a calculation where the elementary nucleon-nucleon cross sections
were raised by a factor two.
The results (full squares marked '$2\sigma$'
in the right panels) show an increased transverse
rapidity variance, as one would expect from the increase of the elementary
collision frequency, but also a rise of the heavy cluster multiplicity,
shown for the interval $|y^{(0)}| < 0.5$,
supporting our observation that cluster production is in a direct correlation
with flow. 
The introduction of more copious binary interactions increases the adiabaticity
(and hence the cooling effect) of the expansion in a twofold way:
1) the local equilibration is faster and 2) the expansion is slower (since the
diffusion time is larger).
Clearly, we have found evidence for the 'second process' suggested 
in~\cite{bondorf95} where nucleon-nucleon interactions act as seeds for
self-organization leading to new clusters.

A key question is whether cluster production is sensitive to 
relevant features in the
nuclear phase diagram, in particular to the existence and location of
a critical point and of 
liquid-vapor coexistence curves.
When using the IQMD code with its two available momentum dependent
Equations of State, EOS, as input, a so-called 'stiff' one with an
incompressibility  $K=380$ MeV around saturation density, or
alternatively a 'soft' EOS, $K=200$ MeV,
we find sensitivity of cluster production.
While all calculations presented so far in Fig.~\ref{f:fig4} were done
with the soft EOS, one calculation, again for Au+Au, was performed assuming
a stiff EOS: see the open squares (marked 'HM')
in the right panels of Fig.~\ref{f:fig4}.
The variance $var(y_x^{(0)})$ is little changed, i.e. it is caused primarily by
the nucleon-nucleon collision frequency which is only indirectly affected by
switching to another (mean field) EOS.
Cluster production in the interval $|y^{(0)}| < 0.5$, however, 
is lowered when assuming a stiff EOS.

It is too early to draw specific conclusions from this potentially interesting
finding.
What is missing, so far, is the underlying phase diagram implied by
IQMD in its various options.
Recently it was shown~\cite{furuta03} with codes
based on Antisymmetrized Molecular Dynamics
(AMD) that caloric curves can be predicted in {\em controlled} scenarios
(fixed volume or pressure).
This presents an important and necessary link between microscopic models for
equilibrium thermodynamics (and nuclear structure as well) and the complex
dynamic situation found in heavy ion collisions.

In summary, 
cluster production has been  investigated
in central collisions of Ca+Ca, Ni+Ni, $^{96}$Zr+$^{96}$Zr,
$^{96}$Ru+$^{96}$Ru, Xe+CsI and Au+Au
reactions at  $0.4A~{\rm GeV}$ incident energy.
We find that the multiplicity of  clusters with charge $Z \ge 3$,
when reduced to the same number of available charges, grows linearly
with system size and is associated with a mid-rapidity source
with increasing transverse velocity fluctuations.
An increase by about a factor of 5.5 is observed
in the mid-rapidity region
between the lightest system (Ca+Ca) and the heaviest one (Au+Au).
The results, as well as simulations using Quantum Molecular Dynamics,
suggest a collision process where  droplets are created in an expanding,
gradually cooling, nucleon gas.
Expansion dynamics, collective radial flow
and cluster formation are closely linked resulting
from the {\em combined} action of nucleon-nucleon scatterings and the mean
fields.

Finally,
we note that global stopping and directed sideflow data are 
available for the same systems~\cite{reisdorf03}.
The simultaneous reproduction of both 'repulsive' observables (sideflow) and
 'attractive' observables (radial flow and its effect on the degree of
clusterization) with the {\em same} transport code will be a challenging task.
The reward, hopefuly, will be a more precise mapping of
the nuclear phase diagram, including the
liquid-vapour transition,
than was possible in the past.

This work was supported in part by the French-German agreement
between {\small GSI} and {\small IN2P3/CEA} (project No 97-27).






\begin{thebibliography}{99}

 
\bibitem{Heer93}W. A. de Heer,
                Rev. Mod. Phys. 65 (1993) 611.

\bibitem{pochodzalla95}J. Pochodzalla, et al.,
                       Phys. Rev. Lett. 75 (1995) 1040.

\bibitem{natowitz02}J. Natowitz, et al.,
                    Phys. Rev. Lett. 89 (2002) 212701.

\bibitem{chomaz02}P. Chomaz, F. Gulminelli,
                  Prog. Theo. Phys. Suppl. 146 (2002) 135.

\bibitem{borderie02}B. Borderie,
                   J. Phys. G 28 R217 (2002).

\bibitem{dagostino00}M. D'Agostino, et al.,
                     Phys. Lett. B 473 (2000) 219.

\bibitem{moretto02}L. Moretto, et al.,
                   Phys. Rev. C  66 (2002) 041601(R).

\bibitem{elliott02}J.B. Elliott, et al.,
                   Phys. Rev. Lett. 88 (2002) 042701.

\bibitem{kleine02}M. Kleine Berkenbusch, et al.,
                    Phys. Rev. Lett. 88 (2002) 022701. 

\bibitem{karnaukhov03}V.A. Karnaukhov, et al.,
                      Phys. Rev. C 67 (2003) 011601.

\bibitem{williams97} C. Williams,  et al., 
         Phys. Rev. C 55 (1997) 2132;
         Phys. Rev. C 59 (1999) 552.  

\bibitem{borderie01}B. Borderie, et al.,
                   Phys. Rev. Lett. 86 (2001) 3252.   

\bibitem{bondorf95} J.P. Bondorf, D.~Idier and I.N.~Mishustin,   
                  Phys. Lett. B 359 (1995) 261.

\bibitem{schuettauf96}A. Sch\"uttauf, et al., 
        Nucl. Phys. A 607 (1996) 457.

\bibitem{beaulieu96} L. Beaulieu,  et al.,
         Phys. Rev. C 54 (1996) 973.

\bibitem{reisdorf03}W. Reisdorf, et al,
         submitted for publication.                         

\bibitem{kunde95}G.J. Kunde, et al.,
                   Phys. Rev. Lett. 74 (1995) 38.   

\bibitem{chikazumi01}Sh. Chikazumi, et al.,
                     Phys. Rev. C 63 (2001) 024602.

\bibitem{heinz99}U. Heinz, B. V. Jacak,       
                 Annu. Rev. Nucl. Part. Sci. 49 (1999) 529.

\bibitem{gobbi93} A. Gobbi, et al,
         Nucl. Inst. Meth. A 324 (1993) 156.

\bibitem{ritman95} J. Ritman, et al,
         Nucl. Phys. B (Proc.Suppl.) 44 (1995) 708.

\bibitem{reisdorf97}W. Reisdorf, et al,
         Nucl. Phys. A 612 (1997) 493.

\bibitem{jeong94}S.C. Jeong, et al,
         Phys. Rev. Lett. 72 (1994) 3468.

\bibitem{deschauen99}B. de~Schauenburg, Ph-D thesis, Strasbourg,
            France, IReS 99-06 (1999). 

\bibitem{reisdorf99} W.~Reisdorf in
         Proc. $7^{th}$ Int. Conf. Clustering Aspects 
         of Nuclear Structure and Dynamics, Rab, June 14-19 (1999),
         World Scientific Publishing Co., 2000, p.323;
         eds. M.Korolija, Z.Basrak and R.Caplar.

\bibitem{petrovici95}M. Petrovici, et al,
         Phys. Rev. Lett. 74 (1995) 5001.

\bibitem{bertsch88}G. F. Bertsch, S. Das Gupta,
         Phys. Rep. 160 (1988) 189.

\bibitem{hartnack98} C.~Hartnack, et al.,
         Eur. Phys. J. A 1 (1998) 151.

\bibitem{reisdorf04} W.~Reisdorf {\em et al.},
         to be published.

\bibitem{aichelin91} J. Aichelin,
         Phys. Rep. 202 (1991) 233.

\bibitem{tsang93}M. B. Tsang, et al,
         Phys. Rev. Lett. 71 (1993) 1502.

\bibitem{hong02}B. Hong, et al.,
                     Phys. Rev. C 66 (2002) 034901.

\bibitem{furuta03}T. Furuta, A. Ono,
          nucl-th/0305050v1


\end{thebibliography}
\end{document}